\documentclass{raa}  
\usepackage{graphicx, times}
\usepackage{natbib}
\usepackage{amssymb,amsmath}
\usepackage{CJKutf8}
\bibpunct{(}{)}{;}{a}{}{,}
\usepackage{diagbox}
\usepackage{appendix}
\usepackage{threeparttable}
\usepackage{xcolor}
\usepackage{subcaption}
\usepackage{url}
\usepackage[a4paper=true,driverfallback=dvipdfm,pagebackref=true]{hyperref}
\hypersetup{colorlinks = true, linkcolor = green, anchorcolor = red, citecolor = blue, filecolor = red, pagecolor = red, urlcolor = red}
\usepackage{threeparttable}
\usepackage{orcidlink}
\mathchardef\mhyphen="2D

\begin{document}

   \title{Toward High-Precision Astrometry with CSST Using Multi-Gaussian Fitting of PSF}

   \volnopage{Vol.0 (20xx) No.0, 000--000}      
   \setcounter{page}{1}          

  \author{Jialu Nie\inst{1,2}, Peng Wei\inst{1}, Zihuang Cao\inst{2,3}, Yibo Yan\inst{1,2}, Chao Liu\inst{1,2,4,5}, Hao Tian\inst{1}, Xin Zhang\inst{1} and Hai-Jun Tian\inst{6}}

    \institute{ Key Laboratory of Space Astronomy and Technology, National Astronomical Observatories, Chinese Academy of Sciences, Beijing  100101, China; {\it niejialu@nao.cas.cn, weipeng01@nao.cas.cn, liuchao@nao.cas.cn}\\
    \and
    University of Chinese Academy of Sciences, Beijing 100049,  China \\
    \and
    Key Laboratory of Optical Astronomy, National Astronomical Observatories, Chinese Academy of Sciences, Beijing 100101, China\\
    \and 
    Institute for Frontiers in Astronomy and Astrophysics, Beijing Normal University, Beijing 102206, China
    \and
    Zhejiang Lab, Hangzhou 311121, China \\
    \and
     Space Information Research Institute, Hangzhou Dianzi University, Hangzhou 310018, China
    }

    \abstract{
The Chinese Space Station Survey Telescope (CSST) presents significant potential for high-precision astrometry. In this study, we show that the point spread function (PSF) modeled by the discrete PSF with Multi-Gaussian function can effectively enhance the astrometric accuracy. We determine that the PSF profile can be accurately modeled by three Gaussians, which takes advantage of reduced computational complexity in PSF convolution. 
In sparse star fields, the lowest centering accuracy we obtain after aberration correction can be below 1 mas. We find that the proper motion errors remain below 1.0 mas/yr for point sources with five observations and approximately 0.8 mas/yr for seven observations with a time baseline of around 3.5 years.
We finally demonstrate that the precision of our position measurements for stars fainter than 21 mag in the simulated CSST crowded field is better than the results from both SExtractor and DOLPHOT. 
    }
   \authorrunning{Nie et al. }  
   \titlerunning{}
   \maketitle

\section{Introduction}
\label{sec:intro}
Astrometry is dedicated to the precise measurement of the positions of celestial bodies and represents one of the oldest fields of study in astronomy \citep{Li_2006, van_Altena_2013}. Its primary objectives include determining stellar positions, proper motions, and parallaxes, crucial in describing its spatial motion \citep{vickers_global_2016} and providing valuable insight into the stellar properties. When astrometric parameters are obtained for a large number of stars, population studies become feasible, along with kinematic and dynamical analyses of the different components of the Milky Way and galaxies. 
In particular, dense star fields, often of high scientific interest and typically consisting of star clusters, dwarf galaxies, and the Galactic bulge, present significant challenges for proper motion calculations due to source blending. This makes distinguishing and extracting stars difficult and increases the measurement uncertainties. Furthermore, the accuracy and stability of the reference frame are crucial, as large systematic biases or significant uncertainties in reference stars can degrade the overall accuracy. Moreover, proper motion calculations usually require multi-epoch data collected over a relatively long time interval. However, variations in observing conditions, field distortions, and systematic errors between epochs may reduce the reliability of long-term measurements, making data alignment more complicated and introduce potential biases. 

 Compared with ground-based equipment, space-borne equipment has a small atmospheric impact, resulting in a high spatial resolution and more stable observing environment. The Hubble Space Telescope (HST) has played a pivotal role in this regard. \cite{zoccali_proper_2001} and \cite{feltzing_new_2002} used HST images to measure the proper motions of the globular clusters NGC 6553 and NGC 6528, respectively, in low-extinction regions within the Galactic bulge. \cite{kuijken_hubble_2002} obtained proper motion data for 36,100 stars in the Baade and Sagittarius I windows. These data facilitated the removal of foreground stars and revealed the rotation of the Galactic bulge. \cite{soto_proper_2014} measured high-precision proper motions for an additional 15,000 stars across three low-extinction fields, achieving an accuracy of 0.9 mas/yr. \cite{shahzamanian_first_2019} conducted further in-depth studies using HST data, publishing proper motion results for 446 stars over a 7-year observational baseline in the highly extinct region of the Galactic center. \cite{clarkson_stellar_2008} utilised HST SWEEPS (Sagittarius Window Eclipsing Extrasolar Planet Search) data, achieving a precision of 0.3 mas/yr in their proper motion measurements. \cite{sohn_m31_2012} were the first to measure the proper motion of the Andromeda galaxy (M31), using HST observations of three stellar fields over a 5- to 7-year baseline, achieving an exceptional precision of 12 \(\mu\)as. In the Gaia era, Gaia provided an unprecedented high-precision reference frame in the optical band. \cite{van_der_marel_first_2016} utilised Gaia DR1 to calculate the proper motions of the Large and Small Magellanic Clouds, investigating the relationship between their proper motions and apparent velocities. Later, \cite{van_der_marel_first_2019} combined the Gaia DR2 data with their 2002 model \citep{van_der_marel_new_2002} to derive the proper motions of M31 and M33. More recently, \cite{bennet_proper_2023} integrated data from both HST and Gaia to compute the proper motions of 12 dwarf galaxies in the Local Group.

The forthcoming Chinese Space Station Survey Telescope \citep[CSST,][]{Zhan2021} has significant potential for astrometry. Its unobstructed 2-meter aperture produces a clean PSF with an FWHM of less than 0.15", enabling a limiting magnitude of 26.3 in the g-band (SNR $>$ 5 for 2 $\times$ 150 s exposures), a performance comparable to that of the HST. With a field-of-view (FOV) of 1.1 square degrees, 300 times larger than the HST, CSST greatly enhances the efficiency of the survey. Its planned sky survey strategy ensures at least two observations for most sky regions over 10 years, with the potential to reach up to 20 observations. These could provide a robust foundation for high-precision proper motion measurements. It is well established that the PSF is essential for high-resolution space data processing, as it reflects the telescope's optical and instrumental response, particularly in densely populated stellar fields. Due to several shared critical features between HST and CSST, such as operation in low-Earth orbit, high spatial resolution, and observation mode, the challenges associated with PSF processing are similar and have a significant impact on the data processing performance, particularly in high-precision astrometric measurements. Critical issues related to the PSF include aberration-induced PSF distortions \citep{Krist1993}, time-dependent PSF variations \citep{bellini_astrometry_2009}, spatial PSF non-uniformity \citep{anderson_psfs_nodate,krist_20_2011}, complex PSF structures \citep{low_high_2004,jee_principal_2007}, and pixelation effects (\citealp{anderson_improved_2003}, Wang et al. in preparation). To address these challenges, the effective PSF \citep[ePSF,][]{anderson_toward_2000} is developed, a refined approach to accurately characterizing the PSF in undersampled images. The ePSF accounts for the pixel response function, modeling how the PSF is distributed over individual detector pixels. This method effectively resolved the pixel phase error (PPE) issue, significantly enhancing the astrometric accuracy of the HST. Beyond its successful application on HST, the ePSF has been widely adopted in other prominent space observation missions and ground-based telescopes, including the James Webb Space Telescope \citep[JWST,][]{nardiello_photometry_2022,nardiello_photometry_2023,libralato_jwst_2024,libralato_high-precision_2024}, the Transiting Exoplanet Survey Satellite \citep{han_tessgaia_2023} and the VISTA Variables in the Via Lactea \citep{minniti_vista_nodate}.

In this paper, we construct empirical PSFs using PSFEx \citep{bertin_automated_2011}. Both Anderson and King's ePSF and our empirical PSF reconstruct the PSF from point-source star images, and the underlying principles are consistent. Focusing on the PSF reconstruction at the center of the FOV, the profiles of our empirical PSF and the ePSF are highly consistent. However, since CSST is a large FOV survey telescope, using the empirical PSF can better characterize the variation of the PSF across a wide FOV compared to the ePSF. Because CSST is designed as an undersampled camera and it will not perform dither observation as HST has done, CSST cannot generate its ePSF with the same level of accuracy as HST.

Inspired by \cite{hogg_replacing_2013}, we found that the PSF of the CSST Survey Camera can be represented in terms of a combination of multiple Gaussian functions with different weights. Combining CSST empirical PSF with a multi-Gaussian model offers advantages in improving PSF characterization and image analysis. Integrating a multi-Gaussian model based on the  empirical PSF allows for precise parameterized modeling of the light distribution, even for complex or asymmetric PSFs. This integration enhances the ability to capture subtle PSF structures that a single Gaussian might miss, leading to improved fitting precision, especially in both the central and outer regions of the PSF, which could reduce biases and improve overall fitting accuracy by addressing optical aberrations, diffraction, and other distortions. Moreover, the multi-Gaussian approach is well suited for deconvolution processes, proving effective in improving image reconstruction quality.

Proper motion and parallax are typically solved together in modern astrometry. However, since parallax has a minor effect on position, this paper focuses on how the new centroid technique impacts the proper motion solution. In Section \ref{sec:PSF}, we first introduce the method for fitting the PSF using multiple Gaussian functions. Section \ref{sec:V} provides the validation of this multi-Gaussian PSF fitting approach. Subsequently, Section \ref{sec:AS} outlines the astrometry process, including stellar centering based on the multi-Gaussian PSF profile and determining proper motion. Section \ref{sec:PMLG} discusses the centering accuracy of stellar point sources in the crowded star field image. Finally, conclusions are presented in Section \ref{sec:Sum}.

\section{Gaussian mixed model of the PSF} 
\label{sec:PSF}
We establish the method of the analytical description of the PSF of CSST in terms of a Gaussian mixed model. The currently available PSF data are from the simulation of CSST. Therefore, we first briefly describe how to simulate the PSF for CSST and reconstruct the empirical PSF. Then we conduct the multi-Gaussian fitting to the empirical PSF. 

\subsection{Reconstruction of empirical PSF}
\label{sec:ePSFR}

\subsubsection{Simulated PSF}
\label{sec:IPSF}
The simulated PSF is one of the important inputs to the CSST image simulation.
The simulated PSF is provided by the optical simulation \citep{Ban2022} using the ray tracing method software Zemax/CODEV. The PSF simulation accounts for various types of static and dynamic conditions. The static PSF simulation is based on the optical model. The mirror fabrication errors, assembly adjustment errors, CCD surface unevenness, 1-$g$ gravity change, and thermal deformation errors are sequentially considered in the simulation. The dynamic errors from jitter and image stabilisation are also taken into account, although they may be over-simplified as a circular Gaussian component.

Because CSST has a wide FOV of more than 1 square degree, the spatial variation of PSF should be carefully considered. To represent the spatial variation of the PSF, \cite{Ban2022} simulated the PSF on a 30$\times$30 grid within the FOV of each CCD. In the meantime, the variation at different wavelengths of the PSF is also taken into account. At each grid point on a CCD, the simulation provides four monochrome PSFs, whose wavelength evenly covers the range of wavelength of the given band, and then sums them up with the weights corresponding to the response curve of the band.
\subsubsection{Simulated image} 
\label{sec:NGP}

To decompose the PSF profile into multiple Gaussians, we first obtain the PSF profile. To mimic the processing of real observations, we reconstruct the empirical PSF from the star profiles in the simulated image, rather than using the simulated PSF directly. The sampling rate of the stars in the simulated image is half the sampling rate of the simulated PSF, and the star image has noise. Thus, the undersampling and noise in the images must be taken into account when reconstructing the PSF.

All simulated images in this work are generated by the CSST simulation software \footnote{\url{https://csst-tb.bao.ac.cn/code/csst_sim/csst-simulation}} (Fang et al. in prep) using the simulated PSF as input. The instrumental effects on the image data include image field distortion, sky background, dark noise, read-out noise (Gaussian), bias, shutter effect, flat-fielding effect, photon response non-uniformity (PRNU) effect, non-linearity, cosmic ray, saturation and spill-out of electron, bad columns, hot pixels, dead (dark) pixels, brighter-fatter effect and charge diffusion effect.
The values of these instrumental effects are specified in Table \ref{tab:SI}.

We select the North Galactic Pole (NGP) area, which means that the central position of the field is at NGP and subsequently the stellar field density and luminosity function of stars are analogous to the NGP, from the CSST Main Survey simulation data. 
The center of the sky region is located at Right Ascension (RA) = 192.8595$^\circ$, Declination (Dec) = 27.1283$^\circ$, and covers an area of about 2.17 deg$^2$. The simulated images continuously cover this area with 135 individual pointings and include 8687 objects, of which 1523 are point sources (stars and quasars). It is worth noting that the current simulations do not account for the temporal variation of PSF. 

\begin{table}[!t]
 \begin{center}
 
 \begin{threeparttable}
 \caption{Basic CSST instrumental parameters used in the image simulations}\label{tab:SI}
 \centering
 \begin{tabular*}{6.5cm}{cccc}
 \hline
    Parameter& & &value\\
 \hline
Gain& & &1.1 $e^{-}$/ADU\\
Pixel scale& & &0.074 arcsec/pixel\\
FOV& & &12' $\times$ 12'\\
Dark currents& & &0.02 $e^{-}$/s/pixel\\
Readout noise& & &5.0 $e^{-}$\\
Exposure time& & &150 s\\
Bias level& & &500 $e^{-}$/pixel\\
Full well\tnote{1}& & &90000 $e^{-}$\\
\hline
\end{tabular*}
\begin{tablenotes}   
\footnotesize           
\item[1] Number of electrons in a pixel full well.
\end{tablenotes}
\end{threeparttable}
\end{center}
\end{table}

The image data generated by the simulation software were then reduced. The bias, dark current, and flat field have been corrected. The fluxes were converted to electrons per second ($e^{-}$/s). The bad pixels, warm/hot pixels, saturated pixels, and cosmic rays were labelled in the mask image (\emph{*\_flg.fits}). The weighted images were calculated to account for photon noise, readout noise, dark current noise and the transfer error of the background correction (\emph{*\_wht.fits}). 
\subsubsection{Empirical PSF}
\label{sec:EPSF}
Based on the reduced images described above, we used SExtractor \citep{bertin_sextractor_1996} for source detection and photometry, then selected a set of bright, isolated and non-saturated stars for the PSF reconstruction.
Before modeling the PSF profile with multi-Gaussians, a non-parametric empirical PSF was reconstructed from the PSFEx output as a linear combination of basis vectors. To select a sample that reconstructs the empirical PSF, we extracted stars in a sparse field of 135 exposures. We used the pixel basis derived from these stamps as the basis vectors and used the super-sampling algorithm along with a simple regularization scheme to model the PSF at 2 times the sampling rate.

The spatial variations of the PSF across the entire chip were also modeled by making the coefficients of the basis vectors themselves a linear combination of the 3rd-order polynomial functions of position in chip coordinates.
The final output of the spatial variant of the empirical PSF is the sum of all basis vectors weighted by their position-dependent polynomial coefficients. The accuracy of the reconstructed empirical PSF is evaluated and discussed in detail in Section \ref{sec:TESTEPSF}.

\subsection{Gaussian mixed PSF model}
\label{sec:MGF}
\begin{figure}[h!]
        \centering
        \includegraphics[scale=0.5]{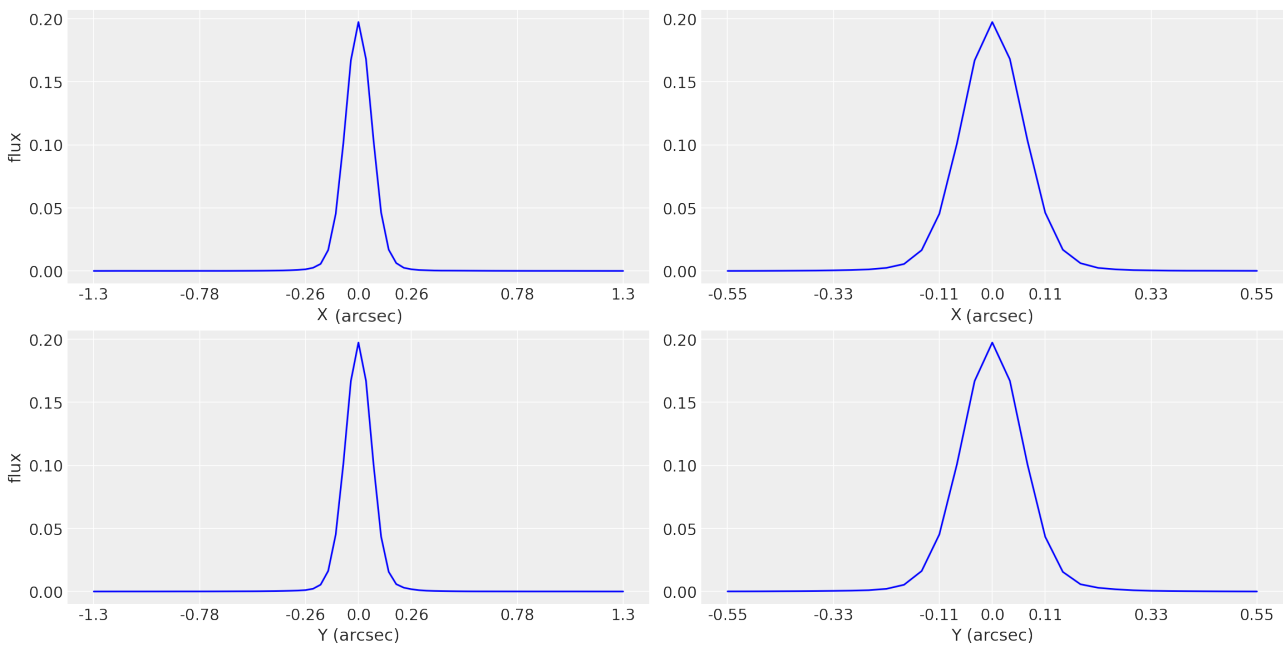}
        \caption{The 2D projected PSF model is illustrated as follows: the left shows the overall profile and the right shows the profile of more than 98\% of the energy left after the cut. The top shows the projection to the X-direction and the bottom shows that to the Y-direction.}
        \label{fig:PSF2D}
        \end{figure}

The reconstructed PSF oversampled image obtained from PSFEx (the left panel of Fig. \ref{fig:PSF2D}) shows that it contains a concentrated core and a broad tail. Given that the tail of the PSF is always very faint and contains a low fraction of the flux, we only consider the PSF profile in the range of a circle centred at the highest flux with a diameter of about 1.1$"$, within which 98\% of the total flux is covered, to fit the PSF with Gaussian mixed model (the right panel of Fig. \ref{fig:PSF2D}). 

PyMC3 is selected for multi-Gaussian fitting for the reconstructed PSF image. It is a Python package for Bayesian statistical modeling and probabilistic machine learning, which is suitable for advanced Markov chain Monte Carlo (MCMC) and variational inference (VI) algorithms \citep{salvatier_probabilistic_2015}. Its flexibility and scalability make it suitable for a large number of problems. 

We first define a Normal likelihood 
    \begin{equation}
        \label{eq2.2a}
            \begin{split}
                P\sim N(\mu,\sigma^2).
            \end{split}
    \end{equation}
The mean, $\mu$, in Eq.~(\ref{eq2.2a}) is the Gaussian mixed PSF model of the following form:
    \begin{equation}
        \label{eq2.2b}
            \begin{split}
                 \mu = \sum\limits_{i}^{K}\omega_i \times G(x,y|\alpha_{x_i},\alpha_{y_i},\beta_{x_i},\beta_{y_i},\gamma_i), 
              \end{split}
        \end{equation}
where $\omega_i$ is the weight of the $i$\,th Gaussian function. 
The G-function is a two-dimensional Gaussian with the image sampled horizontally and vertically for x, y:

\begin{equation}
\label{eqn:4}
\begin{split}
&G(x,y|\alpha_{x_i},\alpha_{y_i},\beta_{x_i},\beta_{y_i},\gamma_i) = \Big(2\pi\beta_{x_i}\beta_{y_i}\sqrt{1-\gamma^2_{i}}\Big)^{-1}\\
&\quad \times exp\Bigg[-\frac{1}{2(1-\gamma^2_{i})}\Bigg(\frac{(x-\alpha_{x_i})^2}{\beta^2_{x_i}}-\frac{2\gamma_i(x-\alpha_{x_i})(y-\alpha_{y_i})}{\beta_{x_i}\beta_{y_i}}+\frac{(y-\alpha_{y_i})^2}{\beta^2_{y_i}}\Bigg)\Bigg],\\
\end{split}
\end{equation}
where $\alpha,\ \beta,$ and $\gamma$ are the mean, variance and correlation coefficient of the Gaussian mixed model, respectively. $\sigma,\ \omega,\  \alpha,\  \beta$ and $\gamma$ are free parameters. Because PyMC3 needs some (loose) prior distribution for the free parameters so that it can obtain an optimal result during the calculation, we chose the following prior distributions for the free parameters:
    \begin{equation}
        \label{eq2.2b}
            \begin{split}
                 \sigma \sim |N(0,1)|,\\
                 \omega\sim Dir(1),\\      
                 \alpha\sim N(0.55,0.037^2),\\
                 \beta\sim |N(0,1.0)|,\\
                 \gamma\sim N(0,0.9^2),
            \end{split}
        \end{equation}
where the units of $\alpha$ and $\beta$ are in arcsecond. $0.037$ is the pixel scale of the CSST PSF and $0.55$ is the radius of the fitted PSF in arcsecond which is obtained by rough estimation by eyes via Fig. \ref{fig:PSF2D}. $Dir$ stands for the Dirichlet distribution that guarantees that $\omega_i$ is between (0, 1) and $\sum\limits_{i} \omega_i$=1.  

\begin{figure}[h!]
        \centering
        \includegraphics[scale=0.25]{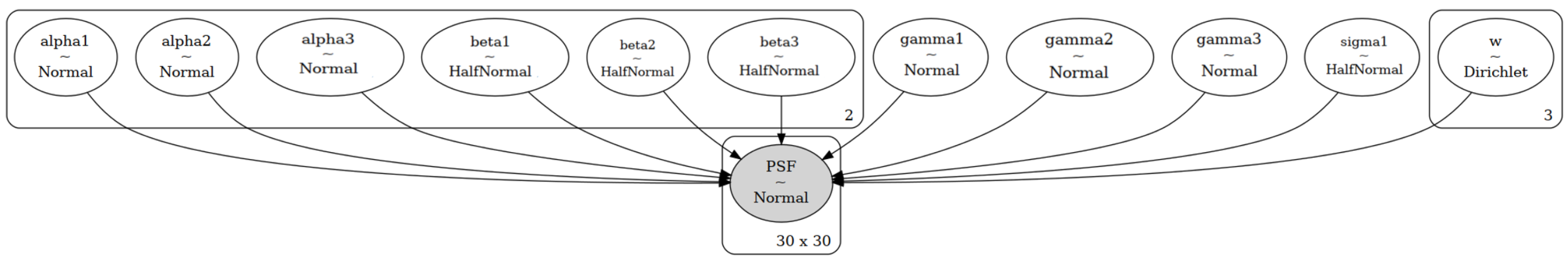}
        \caption{Three-Gaussian mixed model. The upper layer displays the free parameters, in which each ellipse represents a free parameter along with its prior distribution. The number in the box indicates the number of dimensions of the parameter, and the boxes without numbers are one-dimensional. The lower illustrates the prior distribution and dimensionality of the input PSF.}
        \label{fig:model}
        \end{figure}

We used three Gaussian functions to fit the PSF, as shown in Fig. \ref{fig:model}. The first layer displays the free parameters, where the number in each small box indicates the parameter's dimension. Each ellipse features a prior distribution of the parameter. Since we used a 2D Gaussian function for fitting, both $\alpha$ and $\beta$ are 2 dimensions. Additionally, $\omega$ should be a vector with 3 dimensions. 
As mentioned above, the size of the PSF used for fitting is 1.1'' (Fig. \ref{fig:PSF2D}) and the pixel scale of the CSST PSF is about 0.037''/pixel.
So the second layer represents the PSF to be fitted, whose size is 30 × 30 pixel. 

\begin{figure}
  \centering
  \begin{subfigure}[b]{0.8\textwidth}
    \includegraphics[width=\textwidth]{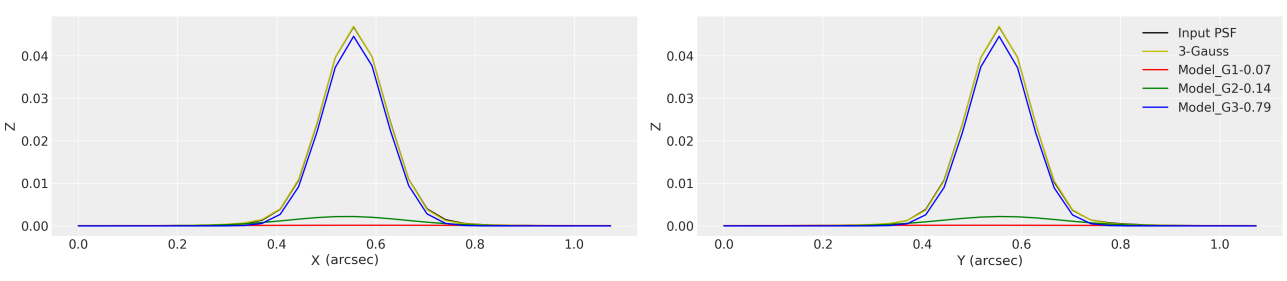}
    \caption{}
    \label{fig:R2}
  \end{subfigure}
  
  \vspace{1em}
  
  \begin{subfigure}[b]{0.8\textwidth} 
    \includegraphics[width=\textwidth]{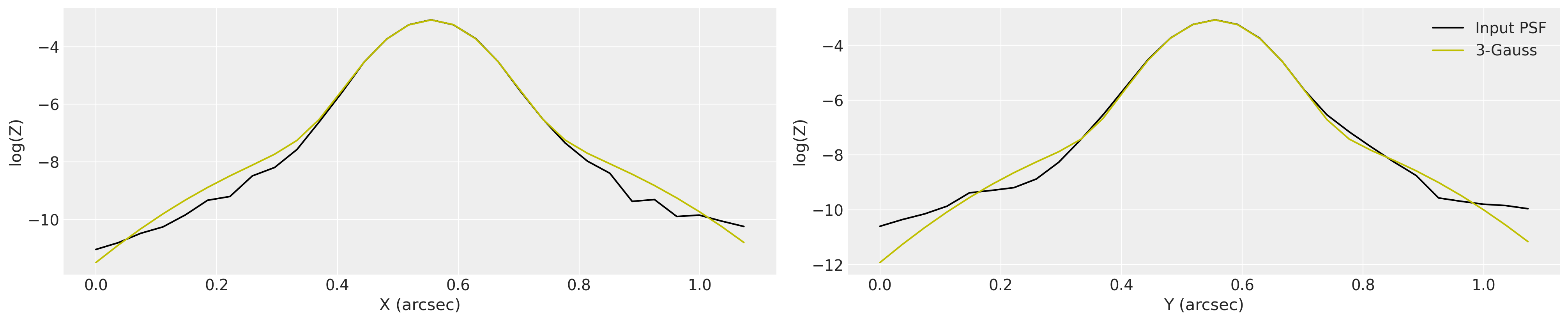}
    \caption{}
    \label{fig:R2Sub}
  \end{subfigure}
  \caption{(a) The best fit three-Gaussian mixed model compared to the input empirical PSF. The red line represents the first Gaussian profile, and the green and blue lines represent the second and third Gaussians, respectively. The fractions of the Gaussians are shown in the legend. The overall profile of the three Gaussians is shown in yellow line and the profile of the input PSF is marked in black. The left panel shows the PSF and the mixed model in the X-projection, while the right panel shows the Y-projection; (b) The difference in the profile of the model and the empirical PSF of the input on a logarithmic scale of normalized flux.}
  \label{fig:main}
\end{figure}

Figure \ref{fig:R2} shows the profile comparison between the fitting result and a selected empirical PSF in the X- and Y-projection. It also displays the contribution of the resulting Gaussian components.
The empirical PSF (black line) has been completely covered by the profile of the three-Gaussian model (yellow line).
A single Gaussian (the blue line) can describe the main central region (80\% of the PSF). The remaining Gaussians describe the wings and the surrounding noise (the remaining 20\% part). To provide a clearer representation, Figure \ref{fig:R2Sub} compares the profiles of the model and the empirical PSF used as input on a logarithmic scale of normalized flux. The posterior results of PyMC3 and the MCMC traces are presented in the Appendix \ref{sec:MCMC}.
Although the more the Gaussians are used the more accurate the mixed model, we finally adopt a three-Gaussian mixed model in this work. We discuss the process of selection of the number of Gaussians in Appendix \ref{sec:NG}.
        
\section{Validation} 
\label{sec:V}

Accurate measurements of the PSF shape are a key requirement in many fields, such as weak gravitational lensing or high-precision astrometry. Therefore, we use the shape indicator used in the weak lensing studies \citep{paulin-henriksson_point_2008,paulin-henriksson_optimal_2009,nie_point_2021} to evaluate the performances of the PSF reconstruction by the empirical PSF and the Gaussian mixed model.

\subsection{Shape Indicator}
\label{sec:AI}
First, we need to evaluate the accuracy of the empirical PSF size $R$ and ellipticity $\epsilon$ at arbitrary positions in the FOV. For this purpose, we judge the difference between the empirical PSF and the Gaussian mixed model related to $R$ and $\epsilon$. Following is the definition of $R$ and $\epsilon$ by \citep{paulin-henriksson_point_2008,paulin-henriksson_optimal_2009,nie_point_2021}
    \begin{equation}
    \label{eq4.1a}
        \begin{split}
            R^2 = Q_{11}+Q_{22},
        \end{split}
    \end{equation}
    \begin{equation}
    \label{eq4.1b}
        \begin{split}
             \epsilon_1 = \frac{Q_{11}-Q_{22}}{Q_{11}+Q_{22}},
             \epsilon_2 = \frac{2Q_{12}}{Q_{11}+Q_{22}},
             \epsilon = \sqrt{{\epsilon_1}^2+{\epsilon_2}^2}.
        \end{split}
    \end{equation}
Here, $Q$s are the second-order moments of the image pixel values
    \begin{equation}
    \label{eq4.1c}
        \begin{split}
            Q_{ij} = \frac{\int d^2xI(x)W(x)x_ix_j}{\int d^2xI(x)W(x)}
        \end{split}
    \end{equation}
where $W(x)$ is the Gaussian weight function used to suppress noise. $x_i$ and $x_j$ are the coordinates of the $(i,j)$ th pixel relative to the center of mass of the star.

Instead of identifying the region of interest to determine the moment measurement region, we employ an adaptive moment method that fits the source image with a 2D elliptic Gaussian profile \citep{hirata_shear_2003}. The elliptic Gaussian profile is iteratively computed as follows: First, the initial weight function $W(x)$ is estimated to be a positive circular Gaussian profile; Then the weighted moments are computed. These results are used to adjust the size, ellipticity, and orientation angle of the elliptical Gaussian function and recalculate the moments with the new elliptical Gaussian profile as the weight function $W(x)$. This process is repeated until the measured moments converge.

In addition, the relative deviation of PSF size (area) $\delta(R^2)/R^2$ and the ellipticity deviation $\delta\epsilon$  are defined as
\begin{equation}
    \label{eq4.1d}
        \begin{split}
            \frac{\delta R^2}{R^2} = \frac{R^2-R^{'2}}{R^2},
            \delta\epsilon = \epsilon-\epsilon',
        \end{split}
    \end{equation}
where $R$ and $\epsilon$ are measured from the simulated PSF that is treated as the ground truth. $R'$ and $\epsilon'$ are measured from the empirical PSF or Gaussian mixed PSF model. $\sigma(\delta(R^2)/R^2)$ and $\sigma(\delta\epsilon)$ are denoted as the standard deviations of $\delta(R^2)/R^2$ and $\delta\epsilon$ of 30 $\times$ 30 grid point locations on the detector area, respectively.

\subsection{Empirical PSF Assessment}
\label{sec:TESTEPSF}
\begin{figure}[h!]
        \centering
        \includegraphics[scale=0.25]{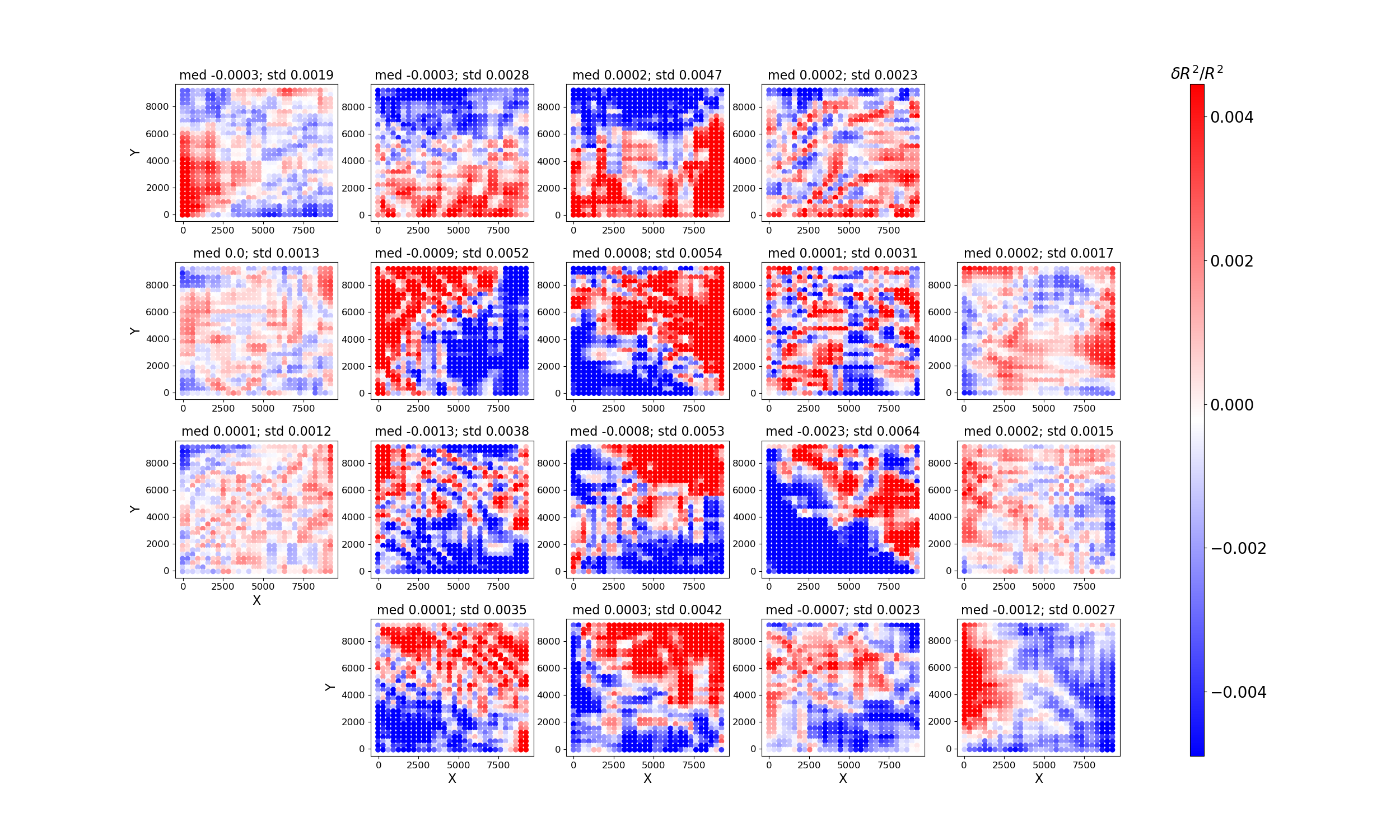}
        \caption{$\delta(R^2)/R^2$ of the empirical PSF on the 18 imaging detectors, each of which contains 30$\times$30 positions, are calculated from the mock images of 135 simulating exposures. The layout of detector subplots aligns with the focal plane detector configuration, with the systematic error (median of $\delta(R^2)/R^2$) and the random error (standard deviation of $\delta(R^2)/R^2$) of the individual CCDs specified in the title. The X and Y coordinates on the subplot denote the positions of the sampling points on the single CCD.}
        \label{fig:R2-3}
    \end{figure}

\begin{figure}[h!]
        \centering
        \includegraphics[scale=0.25]{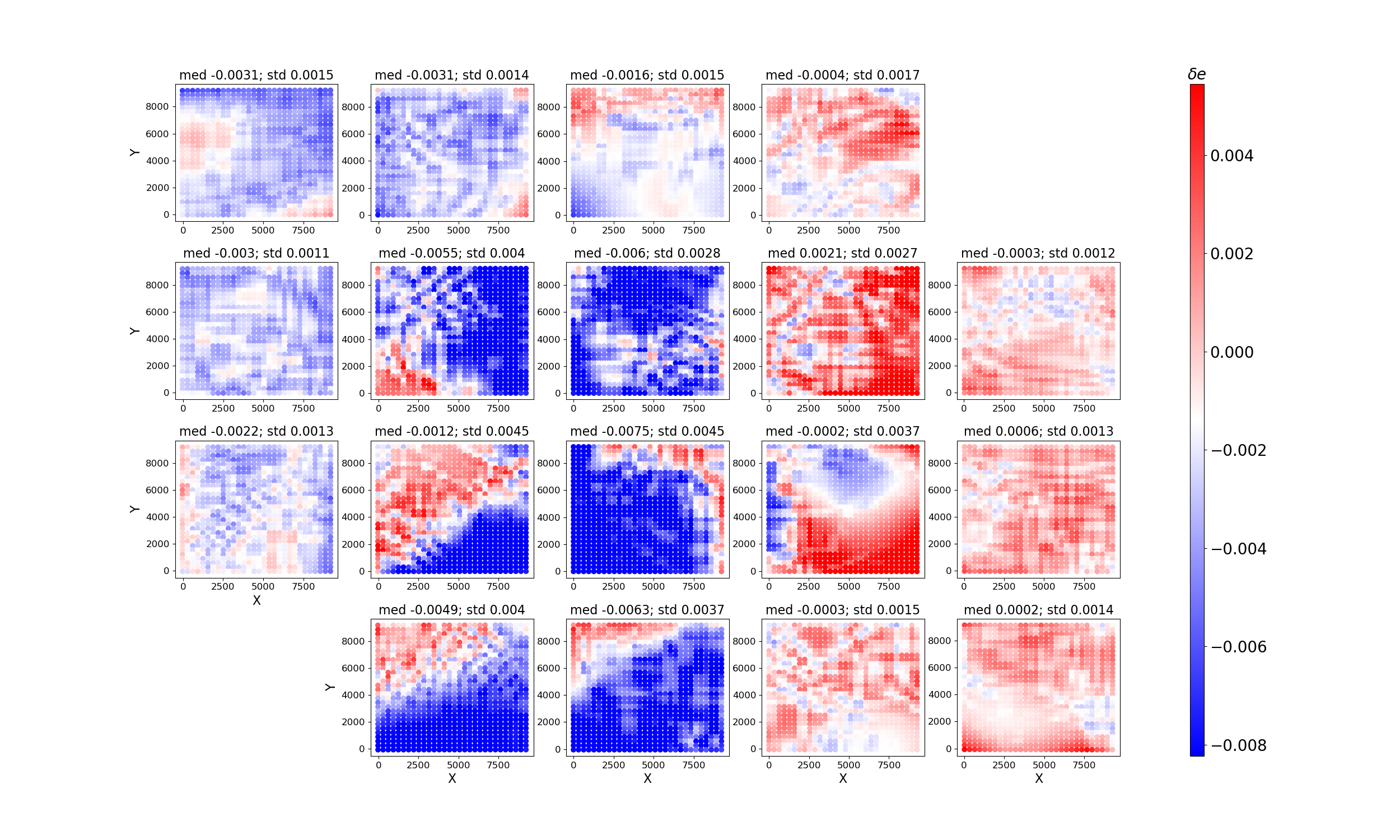}
        \caption{Same as Fig. \ref{fig:R2-3}, but for $\delta\epsilon$.}
        \label{fig:R2-4}
        \end{figure}

According to section \ref{sec:IPSF},  the simulated PSFs are picked out at 30$\times$30 positions per CCD. 
Figures \ref{fig:R2-3} and \ref{fig:R2-4} illustrate the position distributions of $\delta(R^2)/R^2$ and $\delta\epsilon$ of the empirical PSF on the 18 detectors. In some detectors, there are subtle fluctuations visible in the subplots. These fluctuations may arise from the intrinsic small-scale spatial variation of the simulated PSF, which cannot be captured by low-order polynomials in the empirical PSF.
The outcomes presented in Figures \ref{fig:R2-3} and \ref{fig:R2-4} exhibit the satisfactory accuracy of PSF reconstruction.

\subsection{Gaussian mixed PSF model Assessment} 

\begin{figure}[h!]
        \centering
        \includegraphics[scale=0.25]{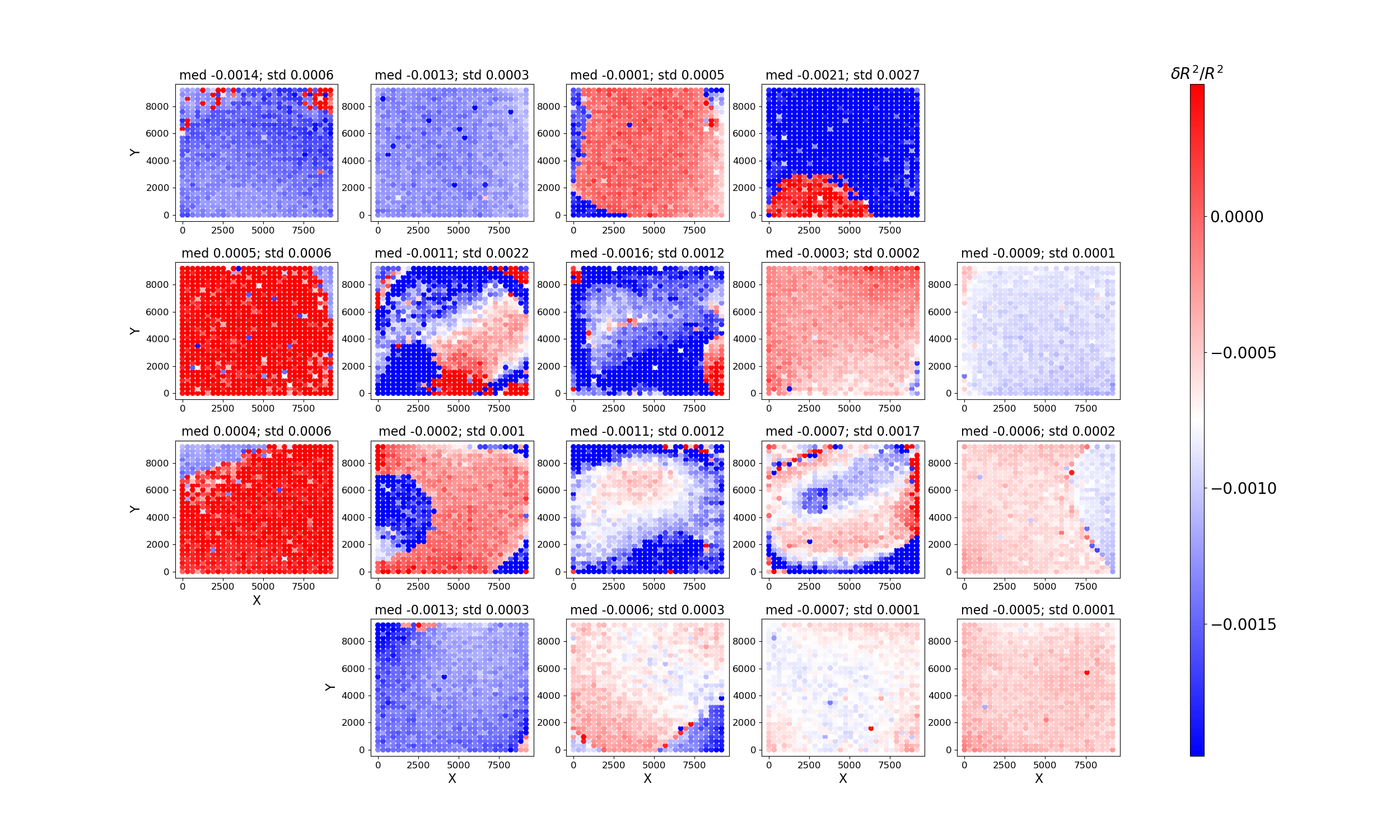}
        \caption{$\delta(R^2)/R^2$ of the three-Gaussian model compared with the empirical PSF. The layout of the subplots and the subplot captions are the same as Fig. \ref{fig:R2-3}.}
        \label{fig:R2-5}
        \end{figure}

\begin{figure}[h!]
        \centering
        \includegraphics[scale=0.25]{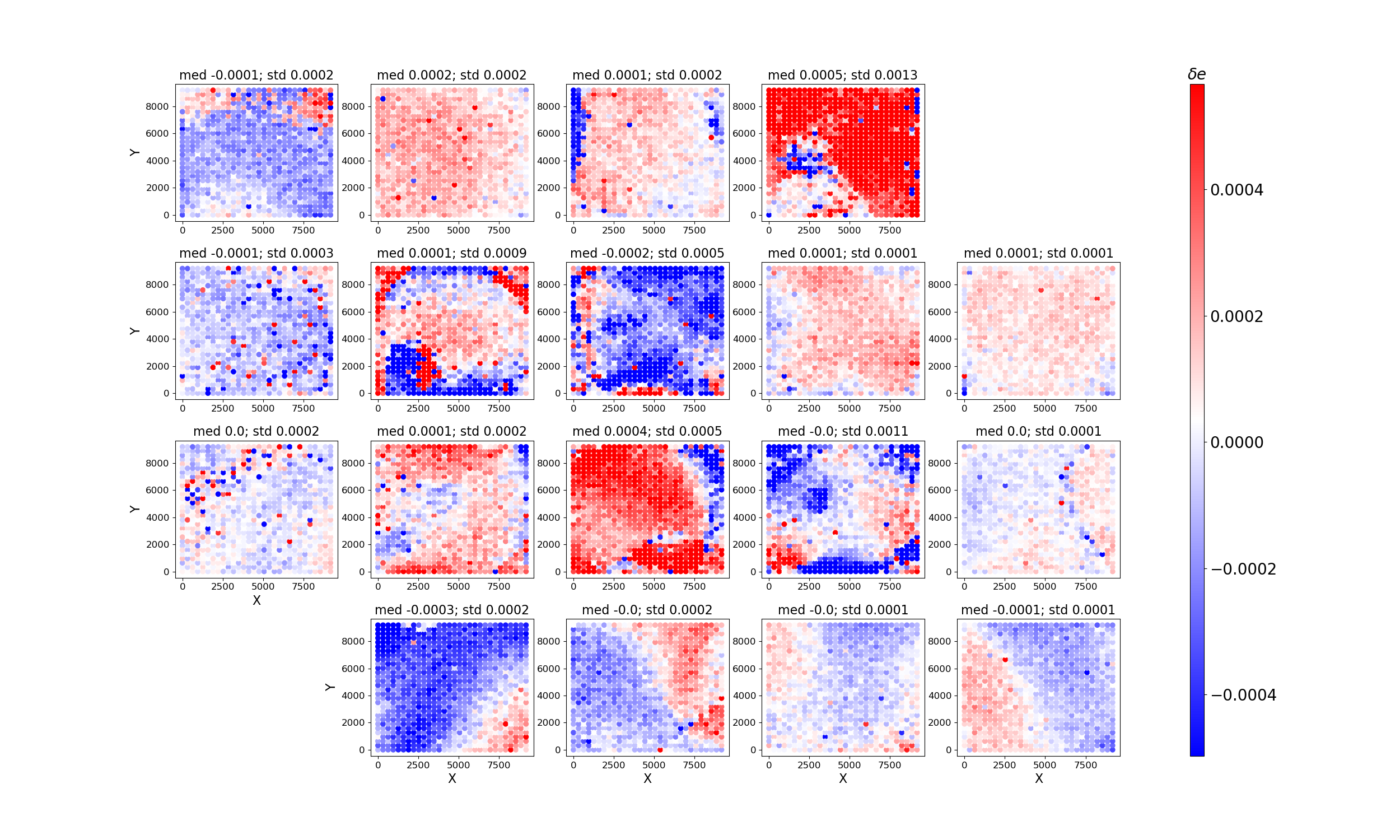}
        \caption{Same as Fig. \ref{fig:R2-5}, but for $\delta\epsilon$.}
        \label{fig:R2-7}
        \end{figure}

We chose to fit the empirical PSF with the Gaussian mixed model rather than to fit directly the simulated PSF because the empirical PSF contains noise and should be the one that we can directly obtain from the observations. As a performance test of the multi-Gaussian mixed model, we fitted the empirical PSF of 30$\times$30 sampling points on each of the 18 CCDs. Figure \ref{fig:R2-5} displays the relative deviation of the radius between the fitting results and the empirical PSF. Fig. \ref{fig:R2-7} is similar but represents errors of ellipticity. Unlike Figures \ref{fig:R2-3} and \ref{fig:R2-4}, the color bar indicates $\delta(R^2)/R^2$ between the best-fit multi-Gaussian model and the empirical PSF, and the subplot captions present the mean and variance of $\delta(R^2)/R^2$ over the 900 points in the corresponding CCD. Fig. \ref{fig:R2-7} is similar but represents $\delta\epsilon$ between the best-fit multi-Gaussian model and the empirical PSF.

By now, we have 30$\times$30 multi-Gaussian PSF models located at fixed grid points on a detector. When we want to apply the multi-Gaussian mixed model to a star located at an arbitrary location on the detector, we need to interpolate the 4 neighbouring PSF models located surrounding the location of the star. A simplified linear interpolation comes up with the linear combination of a total of 12 Gaussians from the 4 neighbouring PSF Gaussian mixed models. By comparing the interpolations of the 4 neighbouring Gaussian mixed models with the empirical PSF, we observe almost the same differences in radius and ellipticity in line with the original sampling point.

We observed specific patterns in Figures \ref{fig:R2-5} and \ref{fig:R2-7}. We analysed that these patterns should come from specific noise in the 135 exposures. We hypothesise that varying noise levels are introduced in the PSF at different sampling points, leading to significant differences between the PSF and Gaussian profiles at specific locations. These patterns may be changed when we change to another sky area.

\section{Stellar astrometry} 
\label{sec:AS}

We  selected multiple observation images from various bands in a simulated sky field in the NGP (more details in Section \ref{sec:NGP}) to investigate the accuracy of astrometry by utilizing a multi-Gaussian PSF model, including point-source centering and proper motion algorithmic performance tests.

\subsection{Data for Astrometric Performance Test}  
\label{sec:SC}

We selected NGP simulation data of the CSST Main Survey simulation with a 3.5-year baseline and one detector-sized overlapping sky region (11.3'$\times$11.3') in seven different epochs with different bands as the test data for stellar centering and proper motion. Observation time and detector information are shown in Table \ref{tab:APM}. 
Note that we do not consider extended sources (galaxies and clusters) in the NGP simulation data.

\begin{figure}[h!]
        \centering
        \includegraphics[scale=0.28]{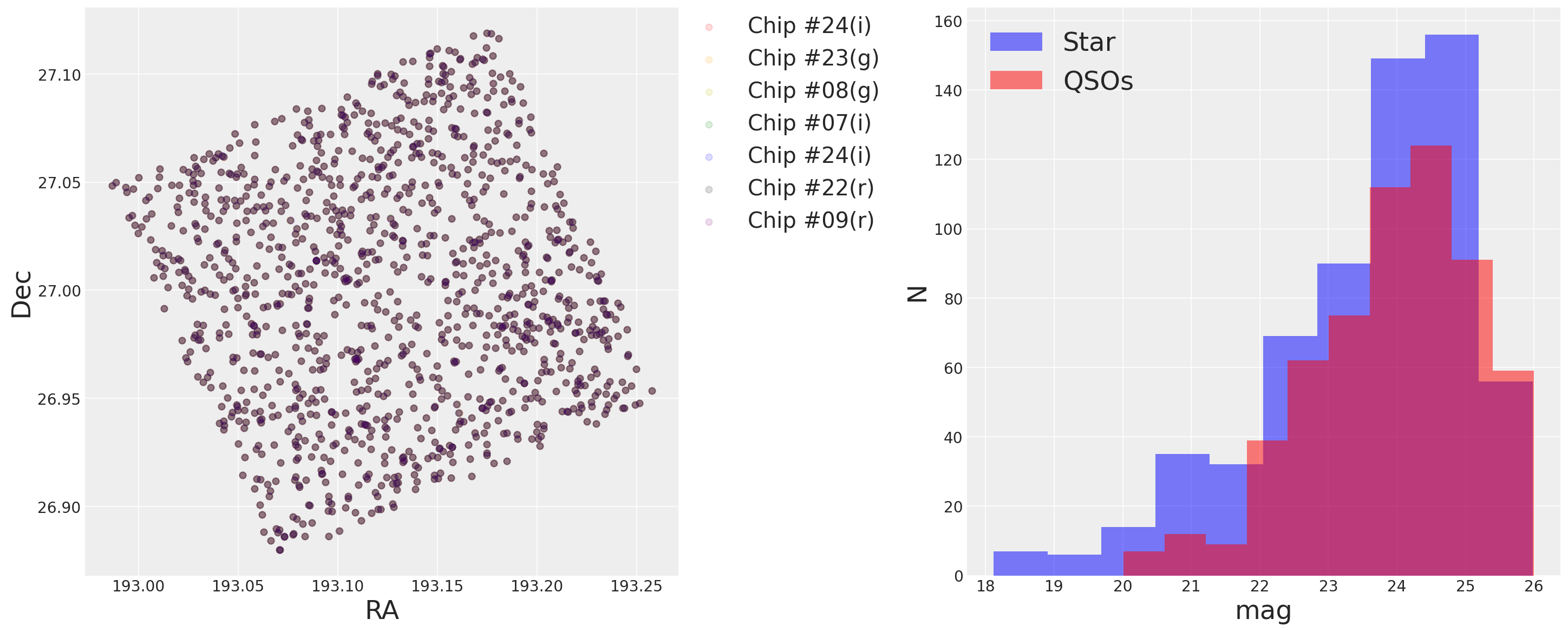}
        \caption{Left panel: Distribution of point sources in the sky area mimic observed at different epochs with different filters. Right panel: The magnitude distribution of the stars (blue) and QSOs (red) to calculate proper motion.}
        \label{fig:A2}
        \end{figure}

The left panel of Fig. \ref{fig:A2} shows the point source distribution of the simulated data on the sky for different detectors at different epochs. Each point is the position of a point source (star or QSO) in equatorial coordinates with color representing different observational epochs. 
The right panel of Fig. \ref{fig:A2} shows the magnitude distribution of stars and QSOs. The star and QSO number ratio is almost 1:1 (614 stars and 590 QSOs). Most of the stars are fainter than 18 mag and most of the QSOs are fainter than 20 mag.

\subsection{Astrometric algorithms}
\label{sec:SCA}

After obtaining the resulting multi-Gaussian models that best fit the empirical PSFs (see Section \ref{sec:MGF} for details), we use them to determine the stellar center position. First, the simulated images were corrected for bias, flat, and dark current before background subtraction. Then, we used SExtractor to subtract the background and detect the sources. 

Then, we fit the point source image by minimizing the cost function $E$
    \begin{equation}
    \label{eq2.3a}
        \begin{split}
            E = \sum\limits_{i}^{N}\frac{[f_{o}(X,Y)-G(X_c,Y_c,F)]^2}{\sigma^2_i},
        \end{split}
    \end{equation}
where $f_{o}$ is the intensity above the background at the pixel $(X,Y)$, $G$ represents the Gaussian mixed model. 
The center position of the point source, ($X_c,Y_c$), and the maximum flux of the point source, $F$, are unknown parameters.
The variance of flux at the $i$\,th pixel, $\sigma_i$, is composed of three components
\begin{equation}
    \label{eq2.3c}
        \begin{split}
            \sigma^2_i = f_{o}+\sigma^2_{RD}+\sigma^2_{sky},
        \end{split}
    \end{equation}
where $\sigma_{RD}$ and $\sigma_{sky}$ are readout noise and the pixel variance of the background, respectively.

After determining the exact location of the point source for each epoch image, we deploy the following $\chi^2$ to fit the proper motion:
    \begin{equation}
    \label{eq2.3e}
        \begin{split}
            \chi^{2}(a,b) = \sum\limits_{i}^{N}\bigg(\frac{(x'_{i}-x_{I})-(a(t_{i}-t_{I})+b)}{\varepsilon_{i}}\bigg)^2,
        \end{split}
    \end{equation}
here $a$ is the proper motion, $b$ is a term related to parallax. $x_{I}$ and $t_{I}$ are the initial position and initial time, respectively. $x'_{i}$ is represented as
\begin{equation}
    \label{eq2.3f}
        \begin{split}
            x'_{i} = x_{i}-\Delta x_{i},
        \end{split}
    \end{equation}
where $x_{i}$ is the original position of a star at epoch $i$ (i.e. $X_c$ and $Y_c$ in Equation \ref{eq2.3a}). $\Delta x_{i}$ is the average displacement of epoch $i$ derived from the first and $i$\,th epoch.
The position uncertainty $\varepsilon_{i}$ consists of two terms, 
    \begin{equation}
    \label{eq2.3g}
        \begin{split}
            \varepsilon_{i} = \sqrt{\varepsilon^2_{c}+{\varepsilon}^2_{t}},
        \end{split}
    \end{equation}
 where $\varepsilon_{c}$ is the error in determining the center of a star. ${\varepsilon}_{t}$ is described as below:
\begin{equation}
    \label{eq2.3h}
        \begin{split}
            \varepsilon_{t} = \big|\Delta x_{r}-\overline{\mu}_{r}\Delta t \big|,
        \end{split}
    \end{equation}
where $\Delta x_{r}$ is the average displacement of the reference object between the first epoch and other epochs. $\overline{\mu}_{r}$ is the averaged proper motion of the reference objects from the known catalog (e.g. Gaia catalog).
To derive the absolute proper motion of an individual star, we added the proper motions of the reference frame to the relative proper motions:
    \begin{equation}
        \label{eq2.3I}
            \begin{split}
                \mu_{m} = \big(\overline{\mu}_{r}+a \big) \pm \varepsilon_{m},
            \end{split}
        \end{equation}
where $\overline{\mu}_{r}$ is the mean value of the proper motions of the reference objects, $a$ is the relative proper motion of the star (as mentioned in Eq. \ref{eq2.3e}), and $\varepsilon_{m}$ is the proper motion error of star.

In general, when the reference system is constructed from galaxies or QSOs, $\overline{\mu}_{r}=0$. When the reference system is constructed by stars, we need to obtain the proper motion of these reference stars.
        
\subsection{Astrometric results} 
\label{sec:SPM}
\begin{table}[!t]
 \begin{center}
 \begin{threeparttable}
 \caption{Observation time and detector information for different epoches}\label{tab:APM}
 \centering
 \begin{tabular*}{6.5cm}{cccc}
 \hline
    Time& & &CCD (band)\\
 \hline
    2024-12-12T08:07:10
    & & &CCD \#24 (i)\\
    2026-05-30T17:27:18
    & & &CCD \#23 (g)\\
    2026-08-11T02:49:27
    & & &CCD \#08 (g)\\
    2026-08-11T05:50:40
    & & &CCD \#07 (i)\\
    2027-08-09T07:50:24
    & & &CCD \#24 (i)\\
    2027-08-09T09:19:18
    & & &CCD \#22 (r)\\
    2028-07-21T20:02:14
    & & &CCD \#09 (r)\\
\hline
\end{tabular*}
\end{threeparttable}
\end{center}
\end{table}

\begin{figure}[h!]
        \centering
        \includegraphics[scale=0.25]{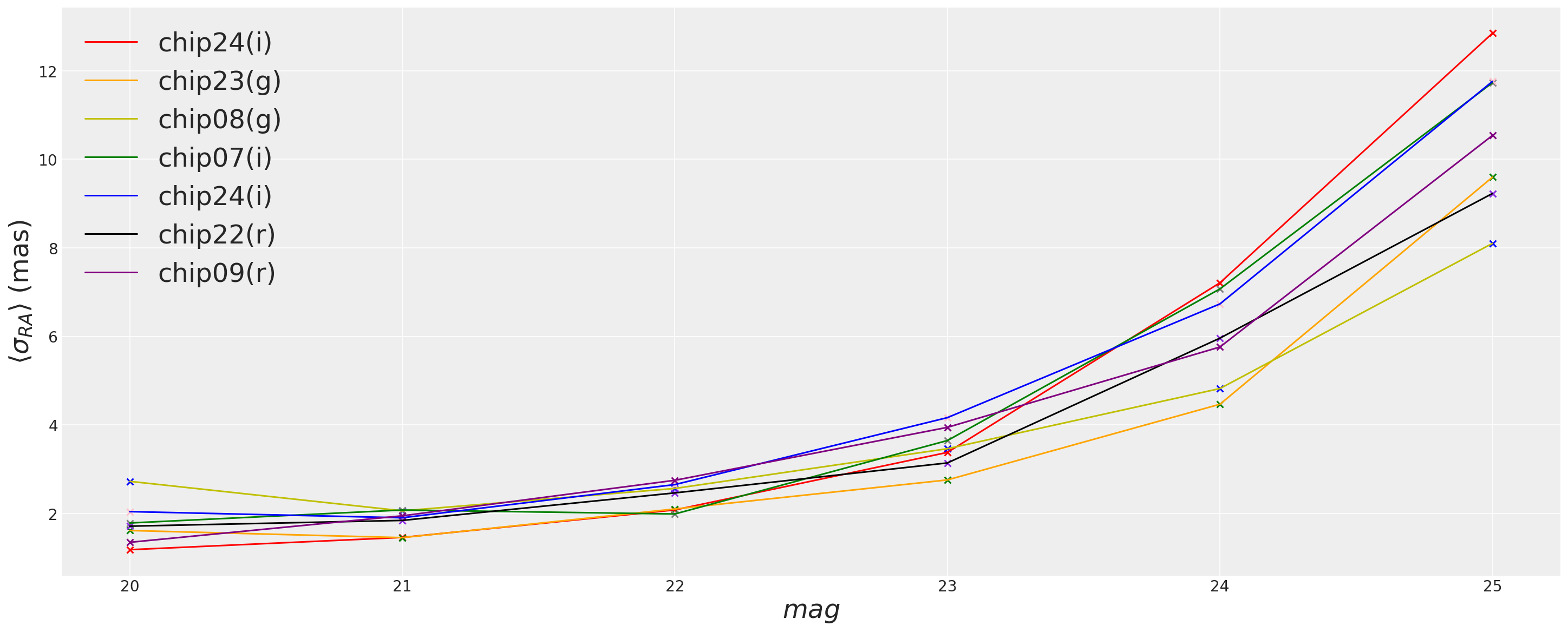}
        \caption{Relationship between position (RA) uncertainty and magnitude for different bands after aberration correction. Different colours represent different bands.}
        \label{fig:7}
        \end{figure}

\begin{figure}[h!]
        \centering
        \includegraphics[scale=0.25]{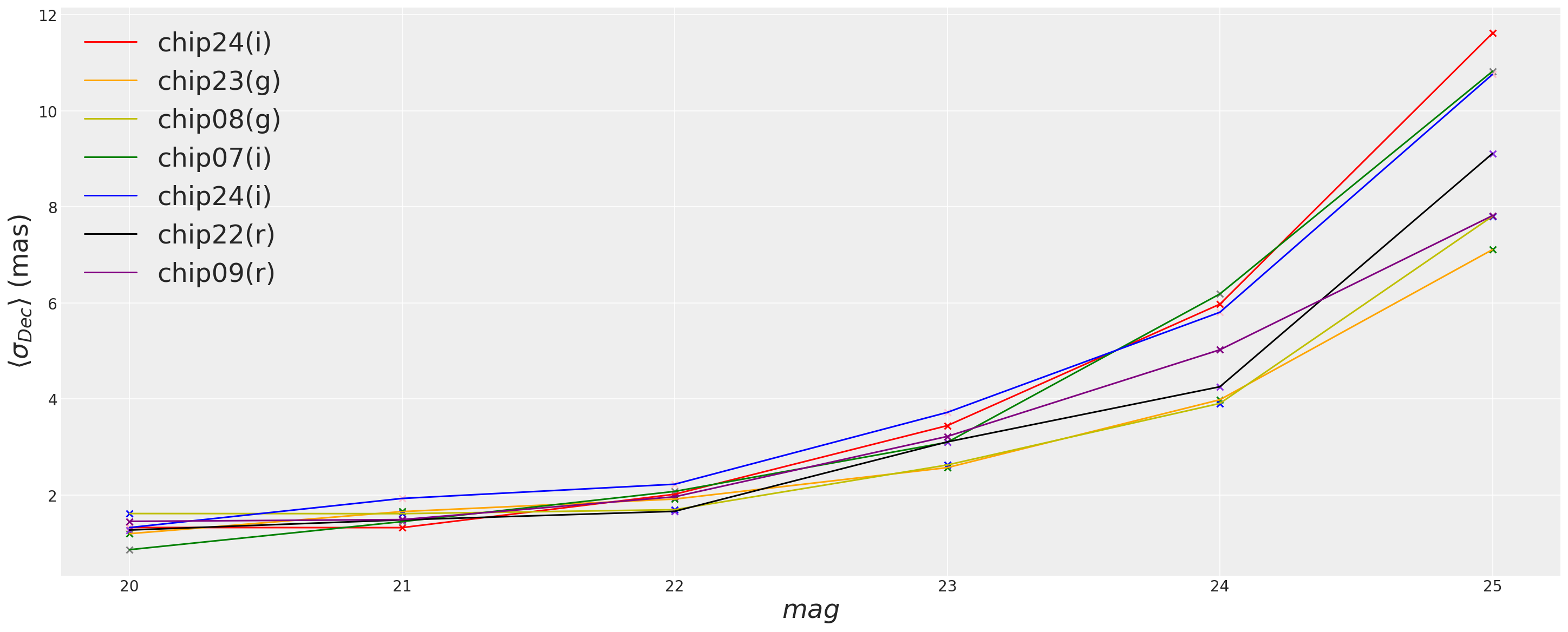}
        \caption{Same as Fig. \ref{fig:7}, but for Dec.}
        \label{fig:8}
        \end{figure}

Instead of using all the QSOs as a reference frame in solving the proper motion, we randomly selected 100 QSOs, and the remaining 490 QSOs were treated as testing point sources and their proper motions were calculated together with the stars. We performed aberration correction after determining the centers of the point sources. The correction of distortion will be discussed in Yan et al. (in prep). The aberration-corrected positional uncertainties for each band are shown in Fig. \ref{fig:7} (RA direction) and Fig. \ref{fig:8} (Dec direction). $\left \langle  \right \rangle$ represents the median value. Under a sparse region like NGP region, the minimum error of the position after aberration correction can be lower than 1 mas (about 0.86 mas).

\begin{figure}[h!]
        \centering
        \includegraphics[scale=0.3]{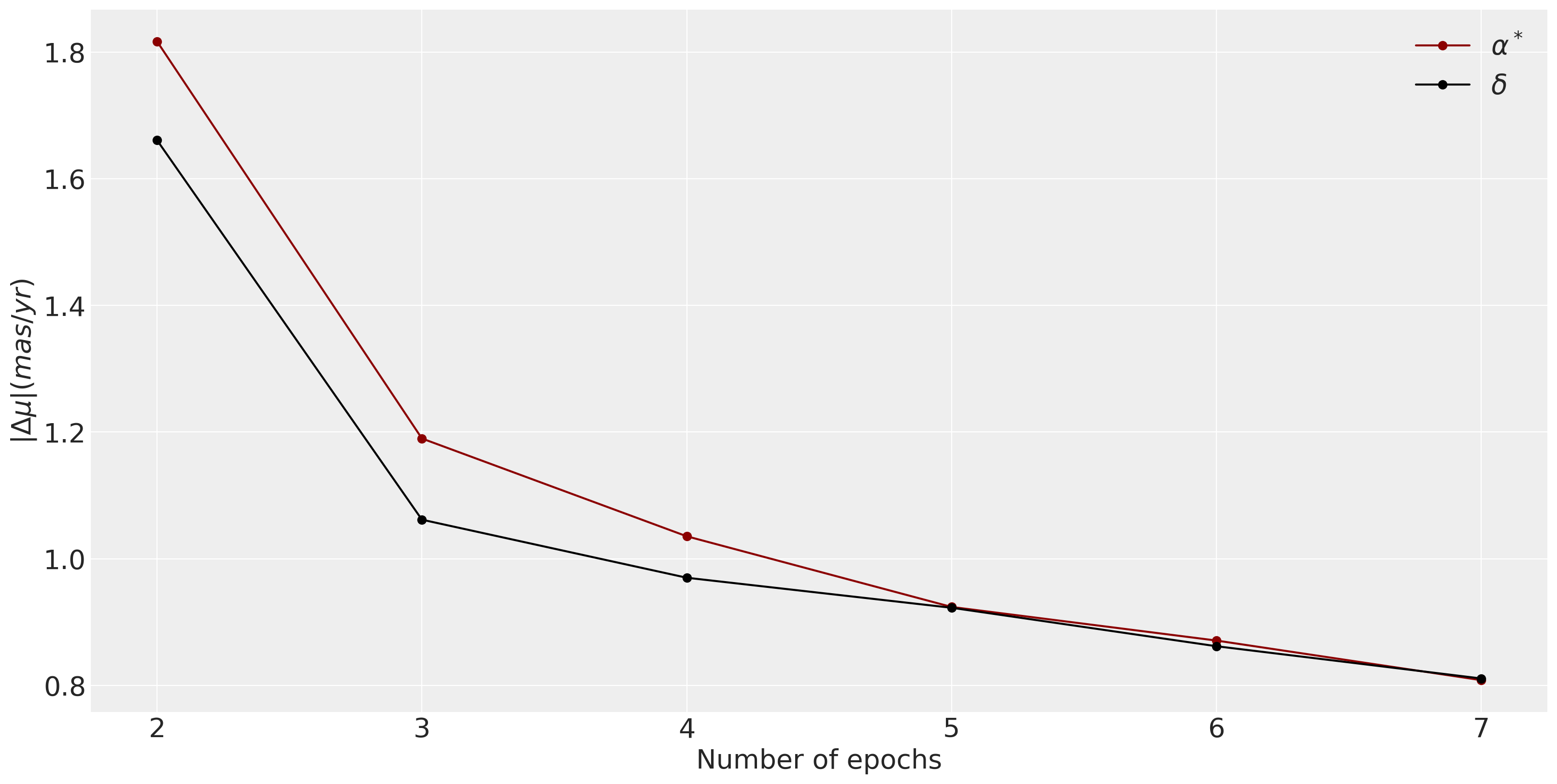}
        \caption{The median of proper motion estimates of the point sources as a function of the number of observations. Red and black lines are for $\alpha^*$ and $\delta$.}
        \label{fig:A2-1}
        \end{figure}

Figure \ref{fig:A2-1} shows the relationship between the proper motion precision of the point sources and the number of observations. Note that $\alpha$ stands for RA, $\delta$ stands for Dec, and $\alpha^*=\alpha \times cos(\delta)$. The observation strategy of the simulation is to make equal time interval observations at the 3.5-year baseline. For example, 2 epochs means that the proper motion calculations only have image data taken at 2024-12-12T08:07:10 and 2028-07-21T20:02:14, and 3 epochs have image data taken at 2024-12-12T08:07:10, 2026-08-11T05:50:40 and 2028-07-21T20:02:14 (see the temporal information in Table \ref{tab:APM}). With a fixed time baseline (3.5 years), a minimum of 5 observations allows the point source to have the proper motion precision of less than 1.0 mas/yr, and the precision of about 0.8 mas/yr after 7 observations.

\begin{figure}[h!]
        \centering
        \includegraphics[scale=0.35]{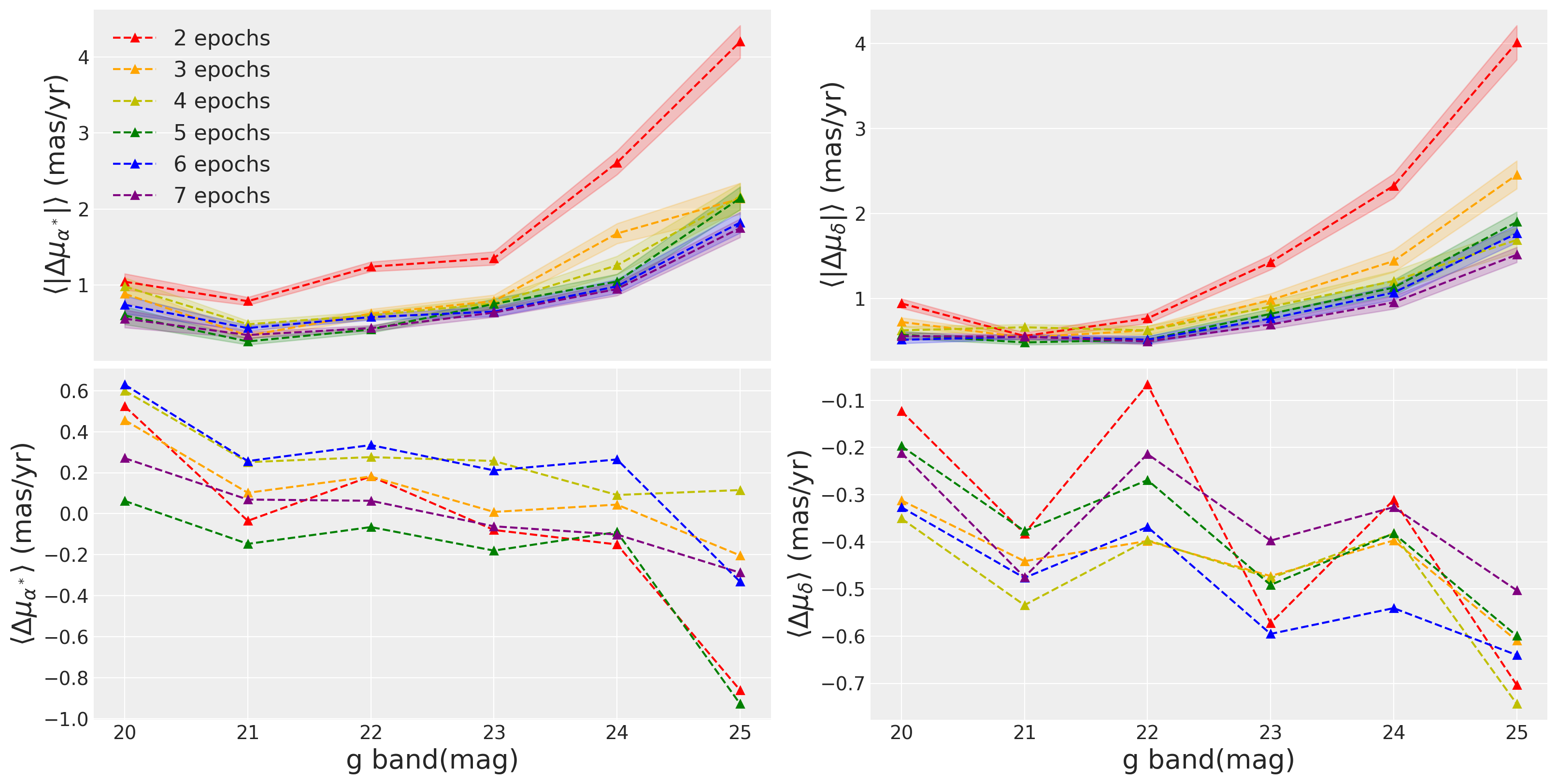}
        \caption{Top panel: The total errors of the proper motion of the QSOs as a function of magnitude. Bottom panel: The systematic errors of the proper motions of the QSOs as a function of magnitude. The different colors represent different numbers of observations under 3.5 years. The left panels show for $\alpha^*$ and the right for $\delta$. }
        \label{fig:A2-2}
        \end{figure}

Figure \ref{fig:A2-2} shows the relationship between the proper motion errors and magnitudes of the QSOs, excluding the stellar results. Similar to the Figure \ref{fig:7}, the quantities with the $\left \langle \right \rangle$ brackets are the median values. The top panels show the total errors (including random and systematic error) and the shaded area is the $1-\sigma$ uncertainty obtained via bootstrapping with 10000 resamplings. The bottom panels show the systematic errors of the two proper motions.

\section{Centering in crowded star field}
\label{sec:PMLG}

In this section, we test our centering method using a Galactic crowded star field as an extreme condition. The field is highly dense and the target of our simulation contains only stars. We use the 3-Gaussian function obtained by fitting the empirical PSF reconstructed in the sparse star field (NGP data) to determine the positions of the stars in the dense star field image.

\begin{figure}[h]
	\centering
	\begin{minipage}{0.5\linewidth}
		\vspace{3pt}
		\centerline{\includegraphics[scale=0.4]{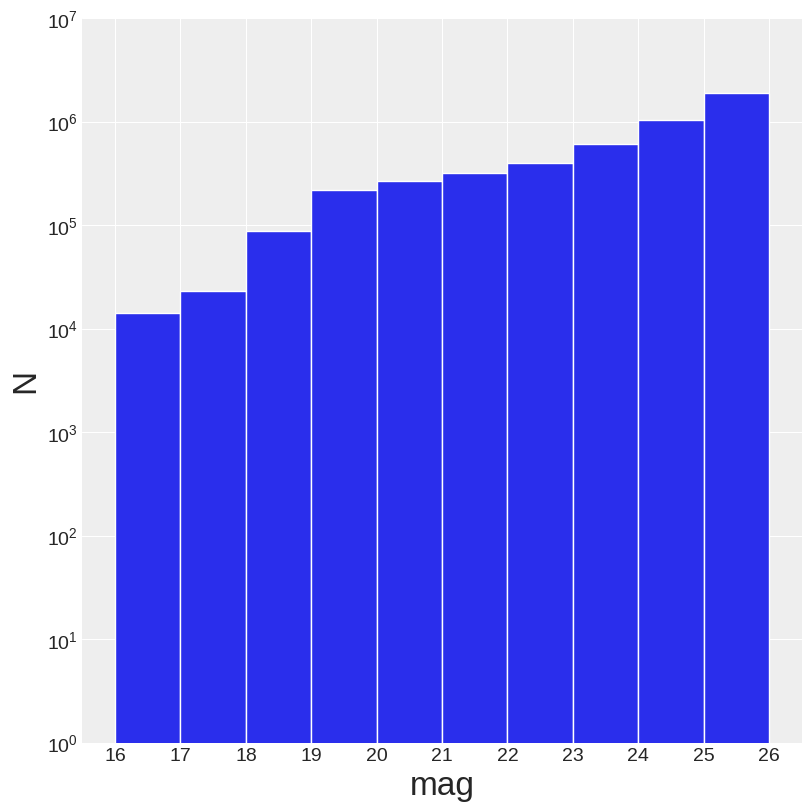}}
	\end{minipage}
	\begin{minipage}{0.5\linewidth}
		\vspace{3pt}
		\centerline{\includegraphics[scale=0.4]{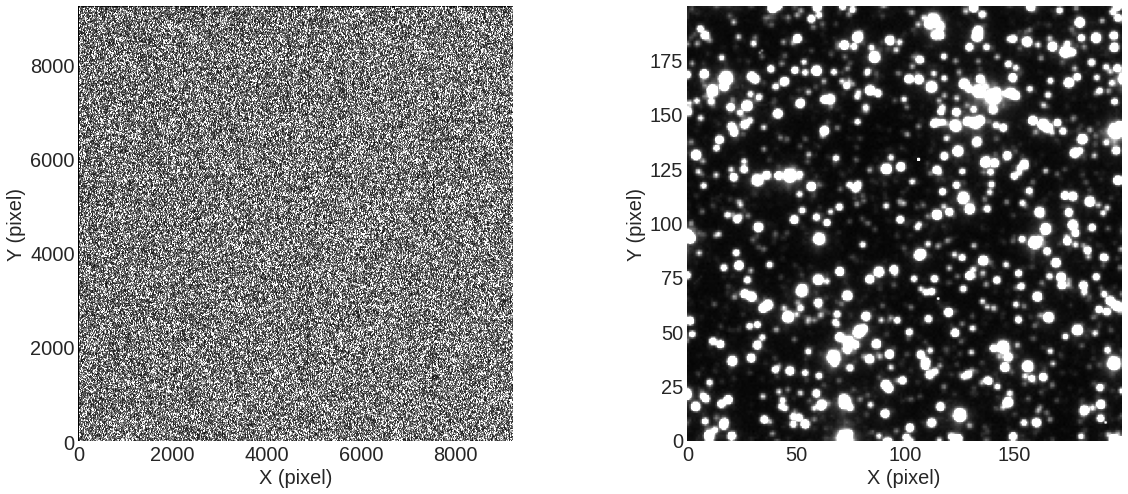}}
	\end{minipage}
	\caption{Top panel: The magnitude distribution of the simulated catalog. Bottom panel: The left is the simulated image of the crowded field from one of the CSST detectors, CCD \#08; the right is an enlarged version of the left. }
	\label{fig4}
\label{fig:S1}
\end{figure}

To generate the simulated catalog of the crowded field in the Galactic bulge, we use Galaxia \citep{sharma_galaxia_2011}, which is a routine for creating synthetic data sets to mimic the Milky Way. We set the colour-magnitude bounds, survey size, and geometry as input parameters to generate a simulated input catalog covering one chip (CCD\,\#08) sky area with a size of 11.3' x 11.3'. The catalog includes about 5.4 million stars that are brighter than 26 in the $g$-band. The catalog contains foreground, bulge, and background stars and is centred at (266$^\circ$, 5$^\circ$) in Ecliptic coordinates. The magnitude distribution histogram of the catalog is shown in the top panel of Figure \ref{fig:S1}.

To get closer to the real conditions of CSST, we selected stars with a magnitude between 16 and 26 mag. This ensures that the simulated image data would be representative of the observations. The simulation image process for the crowded field is similar to that of NGP (see Section \ref{sec:NGP} for details).

The lower left panel of Fig. \ref{fig:S1} shows the result of the crowded field simulated by the CSST simulation software. Note that this is the simulation result of one detector with the size of 9216$\times$9232 pixels. A 200$\times$200 pixel zoom-in image is displayed in the lower right panel of Figure \ref{fig:S1}.

We compare the centering results obtained from our method with SExtractor and DOLPHOT, which are broadly used for the determination of the center position of stars.
We adopt the result of object detection from SExtractor since it provides reasonable completeness and better accuracy than DOLPHOT. Then, we calculate the center positions of the detected stars in the selected detector using 5 different methods: DOLPHOT (calculating centroid coordinates using 2D modified moments), SExtractor\_IMAGE (also calculating centroid coordinates using 2D modified moments), SExtractor\_WIN (calculating centroid coordinates using a window), SExtractor\_PSF (calculating centroid positions by directly fitting a PSF), and our method. It is important to note that the empirical PSF described in Section \ref{sec:EPSF} is employed for the SExtractor\_PSF calculations. The center positions determined by the 5 methods are compared with the corresponding ones in the simulated catalog, which is treated as the ground truth.

\begin{figure}[h!]
        \centering
        \includegraphics[scale=0.3]{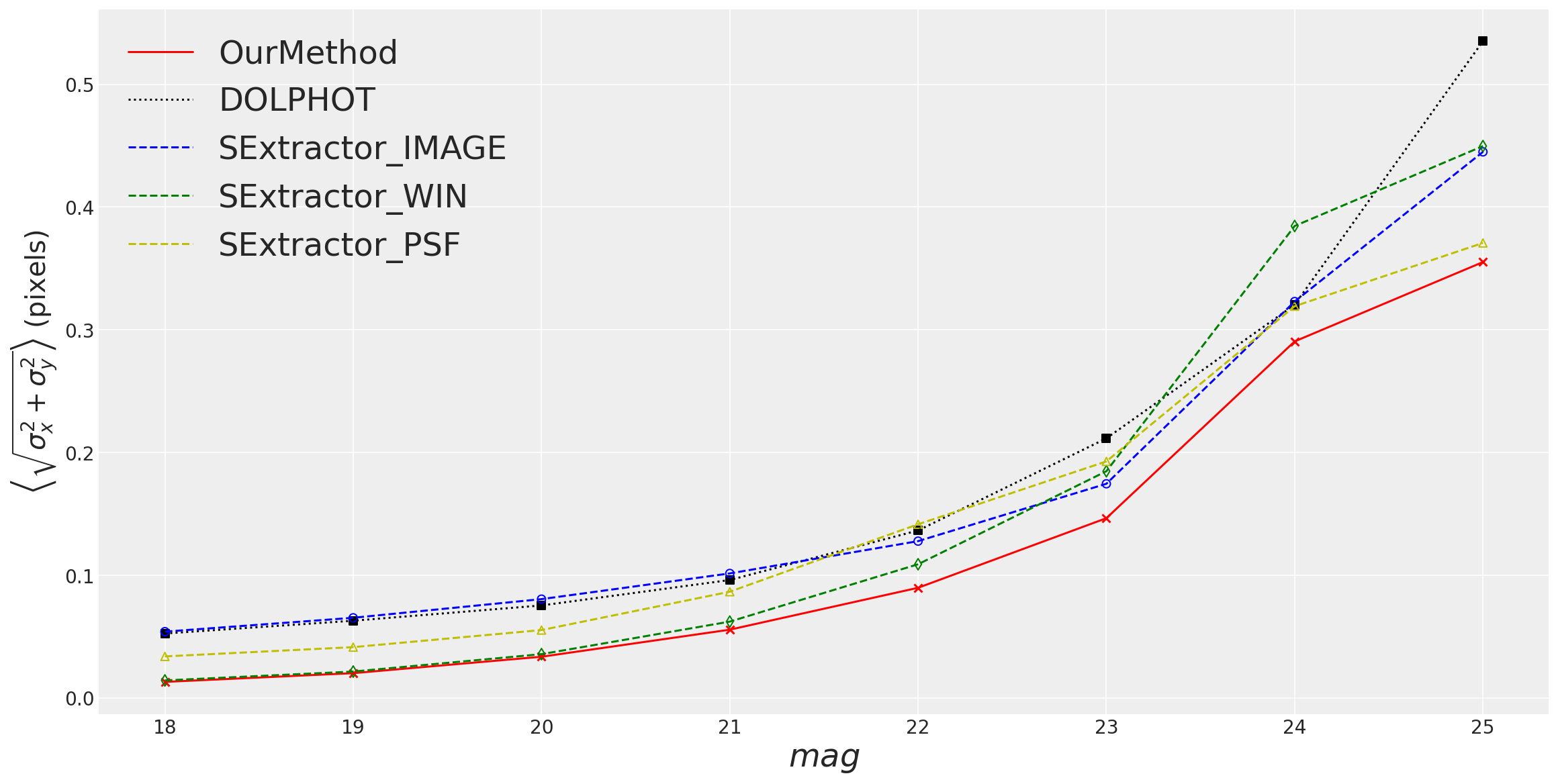}
        \caption{
        Comparison of stellar position errors in different magnitudes between DOLPHOT (black dot), SExtractor and this work (red line). Three different methods of centering provided by SExtractor are shown in blue (SExtractor\_IMAGE), green (SExtractor\_WIN), and yellow (SExtractor\_PSF).
        }
        \label{fig:A1}
        \end{figure}

Fig \ref{fig:A1} displays the median stellar position error as a function of magnitude in the crowded field. In the figure, the errors of the two dimensions ($X$ and $Y$) are merged as $\left \langle \sqrt{\sigma^2_x+\sigma^2_y} \right \rangle$. The step size of the magnitudes is 1\, mag. It is seen that the accuracy of stellar positions obtained by our method is better than that of SExtractor and DOLPHOT for stars fainter than 21\, mag, although SExtractor\_WIN shows similarly good performance for stars brighter than 21\, mag.

\section{Summary}
\label{sec:Sum}

We present the representation of discrete CSST PSFs using continuous combinations of multiple Gaussians.
Firstly, we utilized the CSST Main Survey simulated NGP data and PSFEx to create empirical PSF inputs for fitting.
Subsequently, we employed multiple Gaussian mixed models to fit the empirical PSF. Then we assessed the fitting performance by analyzing deviations in relative radius and ellipticity deviations. Our findings indicate that employing three Gaussians can effectively characterize the empirical PSF.

We applied the multi-Gaussian representation of the PSF to determine the astrometric information  of stars in NGP data. The centering accuracy after distortion correction is preferably better than 1mas. After that, we combined the images from various bands over a time baseline of approximately 3.5 years. Our analysis revealed that 7 observations can yield a proper motion error for point sources of about 0.8 mas/yr. 

Finally, as a discussion, using CSST simulation software and a simulated Galactic bulge catalog, we produced image data to check the centering performance of our three-Gaussian method in the particular star field (crowded dense star field) . We found that our method yielded better accuracy in determining stellar positions compared to SExtractor and DOLPHOT, particularly for stars fainter than 21 mag. It shows that modeling high-precision profiles of the PSFs can significantly enhance astrometric accuracy.

As the CSST is not in operation yet, the data utilized in this study are all simulated data generated by the CSST simulation software, which may exhibit variances from actual observed data. For instance, in real observations, the PSF profile can evolve with the operational duration of the space telescope, introducing complexities that cannot always be accurately simulated.

\begin{acknowledgements}
This work is supported by the China Manned Space Program with grant No. CMS-CSST-2025-A11 and No. CMS-CSST-2021-A08. Z.H.C is supported by the National Natural Science Foundation of China with grant No. 12073047. H.T. is supported by the National Natural Science Foundation of China with grant No. 12103062, 12173046 and the science research grants from the China Manned Space Project. H.J.T is supported by the National Natural Science Foundation of China with grant No. 12373033.
We thank Yuedong Fang and Chengliang Wei for the use of the simulation software and Tianmeng Zhang for the use of the main survey data processing pipeline. 
The sincere thanks go to Dr. Hugh R. A. Jones (Center for Astrophysics, University of Hertfordshire, Hatfield, UK) for his valuable suggestions and revisions to the paper.
\end{acknowledgements}

\bibliographystyle{raa}
\bibliography{ms2024-0314}
\label{lastpage}

\appendix
\section{The results of MCMC}
\label{sec:MCMC}

\begin{figure}[h!]
\centering
\includegraphics[scale=0.4]{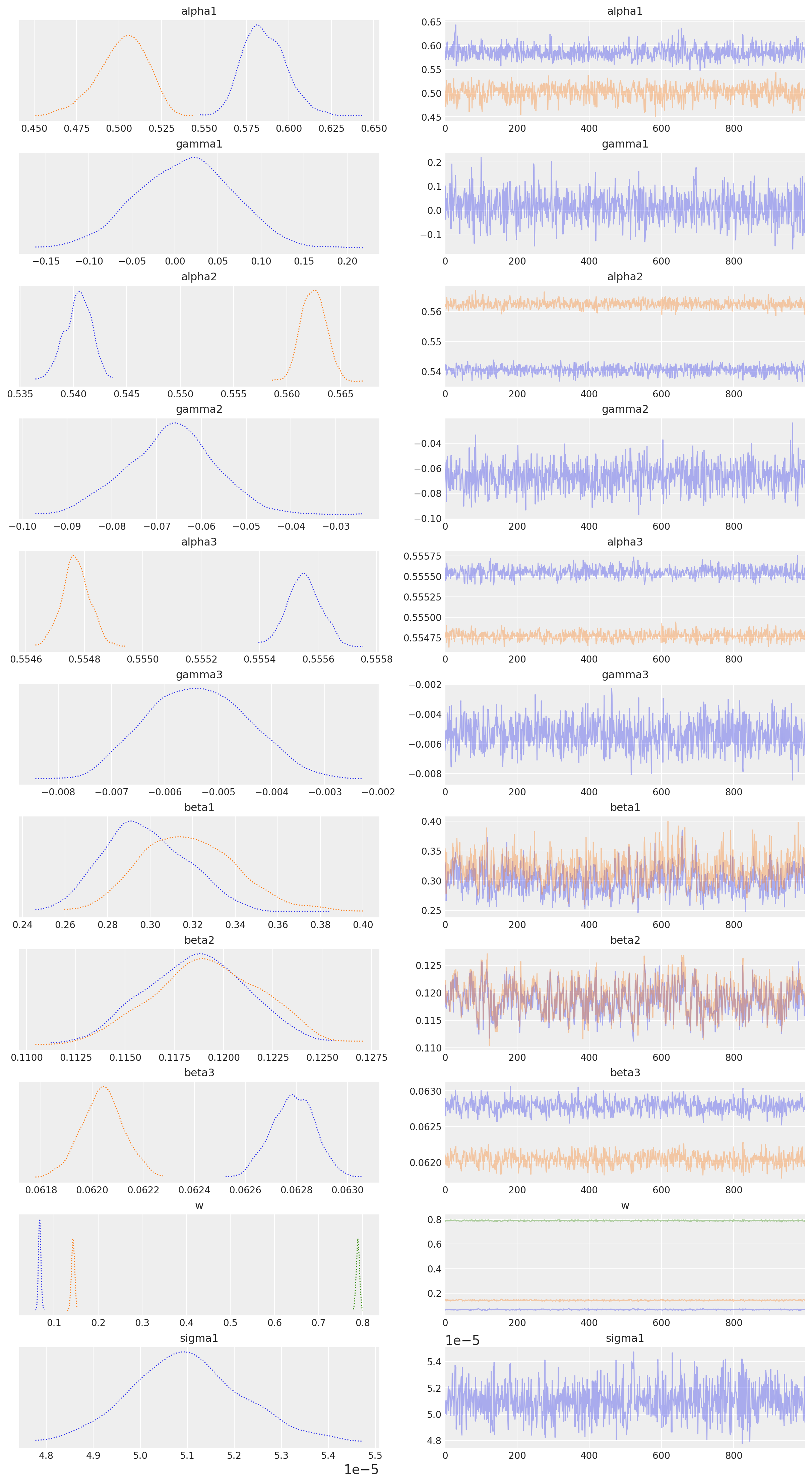}
\caption{Left panel: The resulting distribution of free parameters, blue is the first dimension and orange is
the second; for the weights $\omega$, blue is the weight of the first Gaussian, orange is the weight of the second
Gaussian, and green is the weight of the third Gaussian. Right panel: Trajectories for each dimension of
each parameter.}
\label{fig:MCMC}
\end{figure}

In Fig. \ref{fig:MCMC}, we present the posterior distributions of the free parameters as inferred by PyMC3. The peaks of the normal distributions depicted in the left panel correspond to the optimal parameter values. Meanwhile, the horizontal axis in the right panel indicates the number of MCMC iterations performed.

\section{Number of Gaussian functions}
\label{sec:NG}

\begin{figure}[h!]
        \centering
        \includegraphics[scale=0.18]{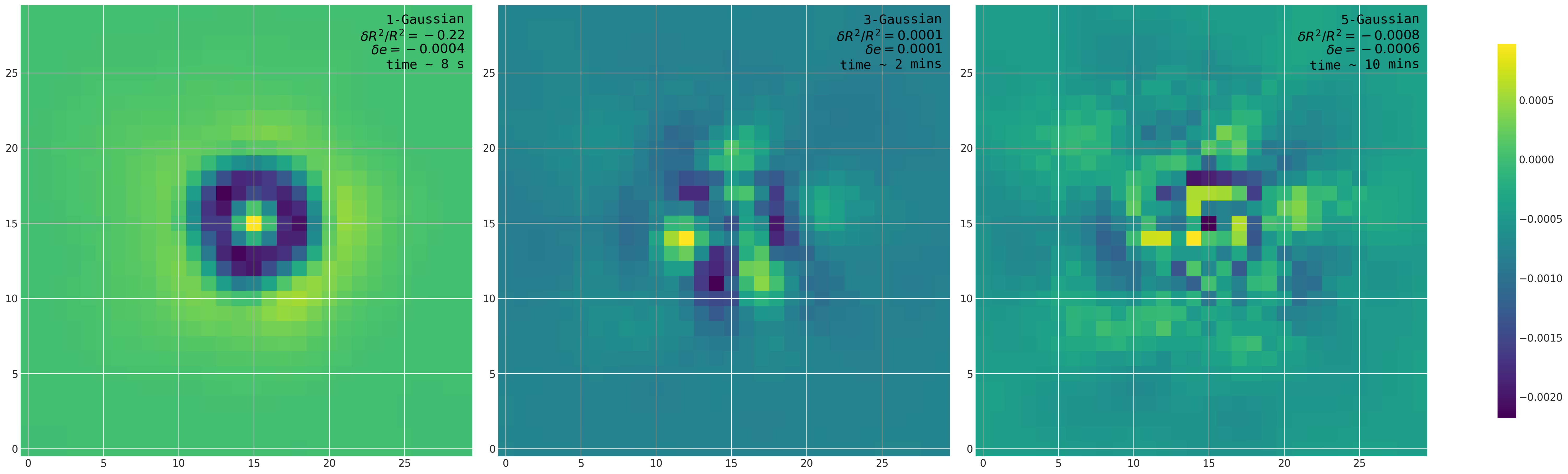}
            \caption{The residuals of the profiles between the different Gaussian mixed models and the empirical PSF,
from left to right panels, 1 Gaussian, 3, and 5 Gaussians, respectively).}
            \label{fig:GT}
            \end{figure}

When evaluating the performance of the multi-Gaussian model, we primarily utilize the criteria outlined in section \ref{sec:AI} (radius and ellipticity) to compare the profiles of the empirical PSF and the best-fit multi-Gaussian model. 
To find the suitable number of Gaussian functions, we fitted the same empirical PSF with 1 Gaussian, 3, and 5 Gaussians, respectively. The empirical PSF and the residual images of the fitted multi-Gaussian models are shown in Fig. \ref{fig:GT}. The \textit{time} in the top right of each panel represents the time taken to fit.

Using only 1 Gaussian function is inadequate for describing the empirical PSF since the profile is not perfectly Gaussian. On the other hand, fitting with 5 Gaussian functions may cause overfitting, despite producing cleaner residual maps. This is because the additional Gaussian functions used to describe the surrounding noise do not fit the profile well in terms of both the radius and ellipticity. About the computation time, fitting with 3 Gaussian functions is acceptable. 
Therefore, we finally chose 3 Gaussians to fit the empirical PSF based on a comparison of the computation time, $\delta(R^2)/R^2$, and $\delta\epsilon$ with other numbers of Gaussians.

\end{document}